\documentclass[11pt]{article}
\usepackage{moriond,epsfig}
\usepackage{amsmath}

\def\be{\begin{equation}}
\def\ee{\end{equation}}
\def\bea{\begin{eqnarray}}
\def\eea{\end{eqnarray}}

\begin{document}
\vspace*{4cm}
\title{CAN NEW GENERATIONS EXPLAIN NEUTRINO MASSES?}

\author{A. APARICI, J. HERRERO-GARCIA, N. RIUS and A. SANTAMARIA}

\address{Depto.de Fisica Teorica, and IFIC, Universidad de Valencia-CSIC \\ Edificio de Institutos de Paterna, Apt. 22085, 46071 Valencia, Spain}

\maketitle\abstracts{
In this talk we explore the possibility that the smallness of the observed neutrino masses is naturally understood in a modified version of the standard model with $N$ extra generations of fermions and $N$ right-handed  neutrinos, in which light neutrino masses are generated at two loops. We find that with $N=1$ it is not possible to fit the observed spectrum of masses and mixings while with $N=2$ it is. Within this extension, we analyse the parameters which are allowed and the possible phenomenological signals of the model in future experiments. Contribution to the proceedings of Les Rencontres de Moriond EW 2011, Young Scientist Forum.}

\section{Introduction}
Neutrino oscillations require at least two massive neutrinos with large mixing, providing one of the strongest evidences of physics beyond the Standard Model (SM). On the other hand, one of the most natural extensions of the SM is the addition of extra sequential generations \cite{Holdom:2009rf}.

LEP II limits on new generation leptons are: $m_{\ell'}>100.8$ GeV and $m_{\nu'}>$ $80.5\:(90.3)\:\mathrm{GeV}$ for pure Majorana (Dirac) particles. When neutrinos have both Dirac and Majorana masses, the bound on the lightest neutrino is $63\,\mathrm{GeV}$. For stable neutrinos LEP I measurement of the invisible $Z$ width, $\Gamma_{\mathrm{inv}}$, implies $m_{\nu'}>39.5\,(45)\,\mathrm{GeV}$ for pure Majorana (Dirac) particles.

The new heavy fermions contribute to the electroweak parameters and might spoil the agreement of the SM with experiment. Global fits of models with additional generations to the electroweak data have been performed and they favour no more than five generations. It should be kept in mind that most of the fits make some simplifying assumptions on the mass spectrum of the new generations and do not consider Majorana neutrino masses for the new generations or the possibility of breaking dynamically the gauge symmetry via the condensation of the new generations' fermions; all these would give additional contributions to the oblique parameters and will modify the fits. Therefore, in view that we will soon see or exclude new generations thanks to the LHC, it is wise to approach this possibility with an open mind.

In this talk (see \cite{Aparici:2011nu} for further details and a complete list of references) we focus on how neutrino masses can be naturally generated at two loops by adding extra families and singlets. Recall that right-handed neutrinos do not have gauge charges and are not needed to cancel anomalies, therefore their number is not linked to the number of generations.

\section{Four generations\label{SM4}}

We extend the SM by adding a complete fourth generation and one right-handed neutrino $\nu_{R}$ with a Majorana mass term \cite{Petcov:1984nz}. We denote the new charged lepton $E$ and the new neutrino $\nu_{E}$. The relevant part of the Lagrangian is

\begin{equation}
\mathcal{L}_{Y}=-\bar{\ell}Y_{e}e_{R}\phi-\bar{\ell}Y_{\nu}\nu_{R}\tilde{\phi}-\frac{1}{2}\,\overline{\nu_{R}^{c}}m_{R}\nu_{R}+\mathrm{H.c.}\ ,\label{eq:4genlag}\end{equation}
where $\ell$ are the left-handed lepton SU(2) doublets, $e_{R}$
the right-handed charged leptons and $\nu_{R}$ the right-handed neutrino. In generation space $\ell$ and $e_{R}$
are organized as four-component column vectors. Thus, $Y_{e}$
is a general, $4\times4$ matrix, $Y_{\nu}$ is a general four-component
column vector with elements $y_{\alpha}$ with $\alpha=e,\mu,\tau,E$,
and $m_{R}$ is a Majorana mass term. 

After spontaneous symmetry breaking (SSB) the mass matrix for the neutral leptons at tree level is a $5\times5$ Majorana symmetric matrix which has only one right-handed neutrino Majorana mass term. Therefore, it leads to two massive Majorana and three massless Weyl neutrinos. Then it is clear that only the linear combination of left-handed neutrinos $\nu_{4}^{\prime}\propto y_{e}\nu_{e}+y_{\mu}\nu_{\mu}+y_{\tau}\nu_{\tau}+y_{E}\nu_{E}$ will pair up with $\nu_{R}$ to acquire a Dirac mass term. Thus, it is convenient to pass from the flavour basis ($\nu_{e},\nu_{\mu},\nu_{\tau},\nu_{E}$) to a new one $\nu_{1}^{\prime},\nu_{2}^{\prime},\nu_{3}^{\prime},\nu_{4}^{\prime}$ where the first three states will be massless at tree level and only $\nu_{4}^{\prime}$ mixes with $\nu_{R}$. 

After this change of basis, $\nu_{\alpha}=\sum_{i}V_{\alpha i}\nu_{i}'$ ($i=1,\text{\ensuremath{\cdots}},4$, $\alpha=e,\mu,\tau,E$) with $V_{\alpha4}\ensuremath=y_{\alpha}/\sqrt{\sum_{\beta}y_{\beta}^{2}}$, we are left with a $2\times2$ mass matrix for $\nu_{4}^{\prime}$ and $\nu_{R}$ which leads to two Majorana neutrinos $\nu_{4}$ and $\nu_{\bar{4}}$ of masses $m_{4,\bar{4}}=\frac{1}{2}\left(\sqrt{m_{R}{}^{2}+4m_{D}^{2}}\mp m_{R}\right)$,where $m_{D}=v\sqrt{\sum_{i}y_{i}^{2}}$, with $v=\langle\phi^{(0)}\rangle$, and $\tan^{2}\theta=m_{4}/m_{\bar{4}}$. If $m_{R}\ll m_{D}$, we have $m_{4}\approx m_{\bar{4}}$ and $\tan\theta\approx1$ (pseudo-Dirac limit) while when $m_{R}\gg m_{D}$, $m_{4}\approx m_{D}^{2}/m_{R}$, $m_{\bar{4}}\approx m_{R}$ and $\tan\theta\approx m_{D}/m_{R}$ (see-saw limit).

\begin{figure}
\centering{}
\psfig{figure=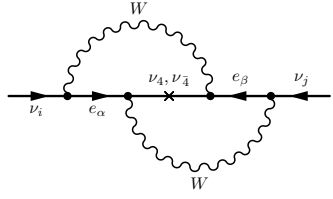,height=1in}
\caption{Two-loop diagram contributing to neutrino masses in the four-generation model. \label{fig:TwoW}}
\end{figure}
Since lepton number is broken by the $\nu_{R}$ Majorana mass term, there is no symmetry which prevents the tree-level massless neutrinos $\nu_{1}^{\prime},\nu_{2}^{\prime},\nu_{3}^{\prime}$ from gaining Majorana masses; they are generated at two loops by the diagram of Figure 1, and are given by

\begin{equation}
M_{ij}=-\frac{g^{4}}{m_{W}^{4}}m_{R}m_{D}^{2}\sum_{\ensuremath{\alpha}}V_{\alpha i}V_{\alpha4}m_{\alpha}^{2}\sum_{\beta}V_{\beta j}V_{\beta4}m_{\beta}^{2}I_{\alpha\beta}\label{eq:masasligeros4gen}\end{equation} where the sums run over the charged leptons $\alpha,\beta=e,\mu,\tau,E$ while $i,j=1,2,3$, and the loop integral $I_{\alpha\beta}$ can be found in \cite{Aparici:2011nu}. It is easy to show that the eigenvalues of the light neutrino mass matrix are proportional to $m_{\mu}^{4},\, m_{\tau}^{4},\, m_{E}^{4}$ which gives a huge hierarchy between neutrino masses:
\begin{equation}
\frac{m_{2}}{m_{3}}\le\frac{1}{4N_{E}^{2}}\left(\frac{m_{\tau}}{m_{E}}\right)^{2}\left(\frac{m_{\tau}}{m_{\bar{4}}}\right)^{2}\le\frac{10^{-7}}{N_{E}^{2}}\ ,\label{eq:ratiomasasligerosapprox}\end{equation}
where we have taken $\ln(m_{\bar{4}}/m_{4})\approx\ln(m_{E}/m_{\bar{4}})\approx1$
and in the last step we used that $m_{E},m_{\bar{4}}\ge100\,\mathrm{GeV}$.
To overcome this huge hierarchy very small values of $N_{E}$ are needed, which would imply that the heavy neutrinos are not mainly $\nu_{E}$ but some combination of the three known ones $\nu_{e},\nu_{\mu},\nu_{\tau}$; however this is not possible since it would yield observable effects in
a variety of processes, like $\pi\rightarrow\mu\nu$, $\pi\rightarrow e\nu$, $\tau\rightarrow e\nu\nu$... This requires that $y_{e,\mu,\tau}\approx10^{-2}y_{E}$, so $N_{E}\approx1$.

Therefore, the simplest version of the model is unable to accommodate the observed spectrum of neutrino masses and mixings. However, notice that whenever a new generation and a right-handed neutrino with Majorana mass at (or below) the TeV scale are added to the SM, the two-loop contribution to neutrino masses is always present and provides an important constraint for this kind of SM extensions.

\section{The five-generation model\label{SM5}}

We add two generations to the SM and two right-handed neutrinos. We
denote the two charged leptons by $E$ and $F$ and the two right-handed
singlets by $\nu_{4R}$ and $\nu_{5R}$. The Lagrangian is exactly
the same we used for four generations but now
$\ell$ and $e$ are organized as five-component column vectors while
$\nu_{R}$ is a two-component column vector containing $\nu_{4R}$
and $\nu_{5R}$. Thus, $Y_{e}$ is a general, $5\times5$ matrix,
$Y_{\nu}$ is a general $5\times2$ matrix and $m_{R}$ is now a general
symmetric $2\times2$ matrix. The model, contrary to the four-generation case, has additional
sources of CP violation in the leptonic sector, however, for simplicity we take all $y_{\alpha}$ and $y_{\alpha}^{\prime}$ real.

As in the four-generation case, the linear combination $\nu_{4}^{\prime}\propto\sum_{\alpha}y_{\alpha}\nu_{\alpha}$
only couples to $\nu_{4R}$ and the combination $\nu_{5}^{\prime}\propto\sum_{\alpha}y_{\alpha}^{\prime}\nu_{\alpha}$
only couples to $\nu_{5R}$. Therefore, the tree-level spectrum will
contain three massless neutrinos (the linear combinations orthogonal
to $\nu_{4}^{\prime}$ and $\nu_{5}^{\prime}$) and four heavy Majorana
neutrinos. For simplicity, we choose $\nu_{4}^{\prime}$ and $\nu_{5}^{\prime}$ orthogonal to each other, i.e., $\sum_{\alpha}y_{\alpha}y_{\alpha}^{\prime}=0$. We change from the flavour fields $\nu_{e},\nu_{\mu},\nu_{\tau},\nu_{E},\nu_{F}$ to a new basis $\nu_{1}^{\prime},\nu_{2}^{\prime},\nu_{3}^{\prime},\nu_{4}^{\prime},\nu_{5}^{\prime}$
where $\nu_{1}^{\prime},\nu_{2}^{\prime},\nu_{3}^{\prime}$
are massless at tree level, so we are free to choose them in any combination of the flavour states as long as they are orthogonal to $\nu_{4}^{\prime}$ and $\nu_{5}^{\prime}$. 

The model should be compatible with the observed universality of fermion couplings and have small rates of lepton flavour violation in the charged sector, which requires $y_{e},y_{\mu},y_{\tau},y_{e}^{\prime},y_{\mu}^{\prime},y_{\tau}^{\prime}\ll y_{E},y_{F},y_{E}^{\prime},y_{F}^{\prime}$. In addition, it should fit the observed pattern of masses and mixings, for instance, reproducing the tribimaximal (TBM) mixing structure. A successful choice of the Yukawas to obtain normal hierarchy (see  \cite{Aparici:2011nu} for an analysis of inverted hierarchy and more details), i.e, $m_{3}\approx\sqrt{\left|\bigtriangleup m_{31}^{2}\right|}\approx0.05$~eV,
$m_{2}\approx\sqrt{\bigtriangleup m_{21}^{2}}\approx0.01$~eV,   will be $y_{\ensuremath{\alpha}}=y_{E}(\epsilon,\epsilon,-\epsilon,1,0)\nonumber$ and $y_{\alpha}^{\prime}=y_{F}^{\prime}(0,\epsilon^{\prime},\epsilon^{\prime},0,1)$
 which, keeping only terms up to order $\epsilon^{2}$, leads to the $5\times5$ unitary matrix  $V$ that passes from one basis to the other, $\nu_{\alpha}=\sum_{i}V_{\alpha i}\nu_{i}^{\prime}$ ($i=1,\text{\ensuremath{\cdots}},5$, $\alpha=e,\mu,\tau,E,F$):
\begin{eqnarray}
V & \approx & \left(\begin{array}{ccccc}
\sqrt{\frac{2}{3}} & \frac{1}{\sqrt{3}}-\frac{\sqrt{3}}{2}\,\epsilon^{2} & 0 & \epsilon & 0\\
-\frac{1}{\sqrt{6}} & \frac{1}{\sqrt{3}}-\frac{\sqrt{3}}{2}\,\epsilon^{2} & \frac{1}{\sqrt{2}}-\frac{1}{\sqrt{2}}\,\epsilon^{\prime2} & \epsilon & \epsilon^{\prime}\\
\frac{1}{\sqrt{6}} & -\frac{1}{\sqrt{3}}+\frac{\sqrt{3}}{2}\,\epsilon^{2} & \frac{1}{\sqrt{2}}-\frac{1}{\sqrt{2}}\,\epsilon^{\prime2} & -\epsilon & \epsilon^{\ensuremath{\prime}}\\
0 & -\epsilon\,\sqrt{3} & 0 & 1-\frac{3}{2}\,\epsilon^{2} & 0\\
0 & 0 & -\epsilon^{\prime}\,\sqrt{2} & 0 & 1-\epsilon^{\prime2}\end{array}\right) \label{eq:rotationNH}\end{eqnarray}
 Assuming that $m_{E,F}\gg m_{4,\bar{4},5,\bar{5}}\gg m_{W}$, we
find: \begin{equation}
m_{2}\approx\frac{3g^{4}}{2(4\pi)^{4}m_{W}^{4}}\epsilon^{2}m_{4D}^{2}m_{4R}m_{E}^{2}\ln\frac{m_{E}}{m_{\bar{4}}}\label{eq:m2-NH}\end{equation}
 \begin{equation}
m_{3}\approx\frac{g^{4}}{(4\pi)^{4}m_{W}^{4}}\epsilon^{\prime2}m_{5D}^{2}m_{5R}m_{F}^{2}\ln\frac{m_{F}}{m_{\bar{5}}}\ ,\label{eq:m3-NH}\end{equation}
and the required ratio $m_{3}/m_{2}\approx5$ can be easily accommodated for different combinations of masses and mixing parameters $\epsilon$, $\epsilon'$.

\begin{figure}
\psfig{figure=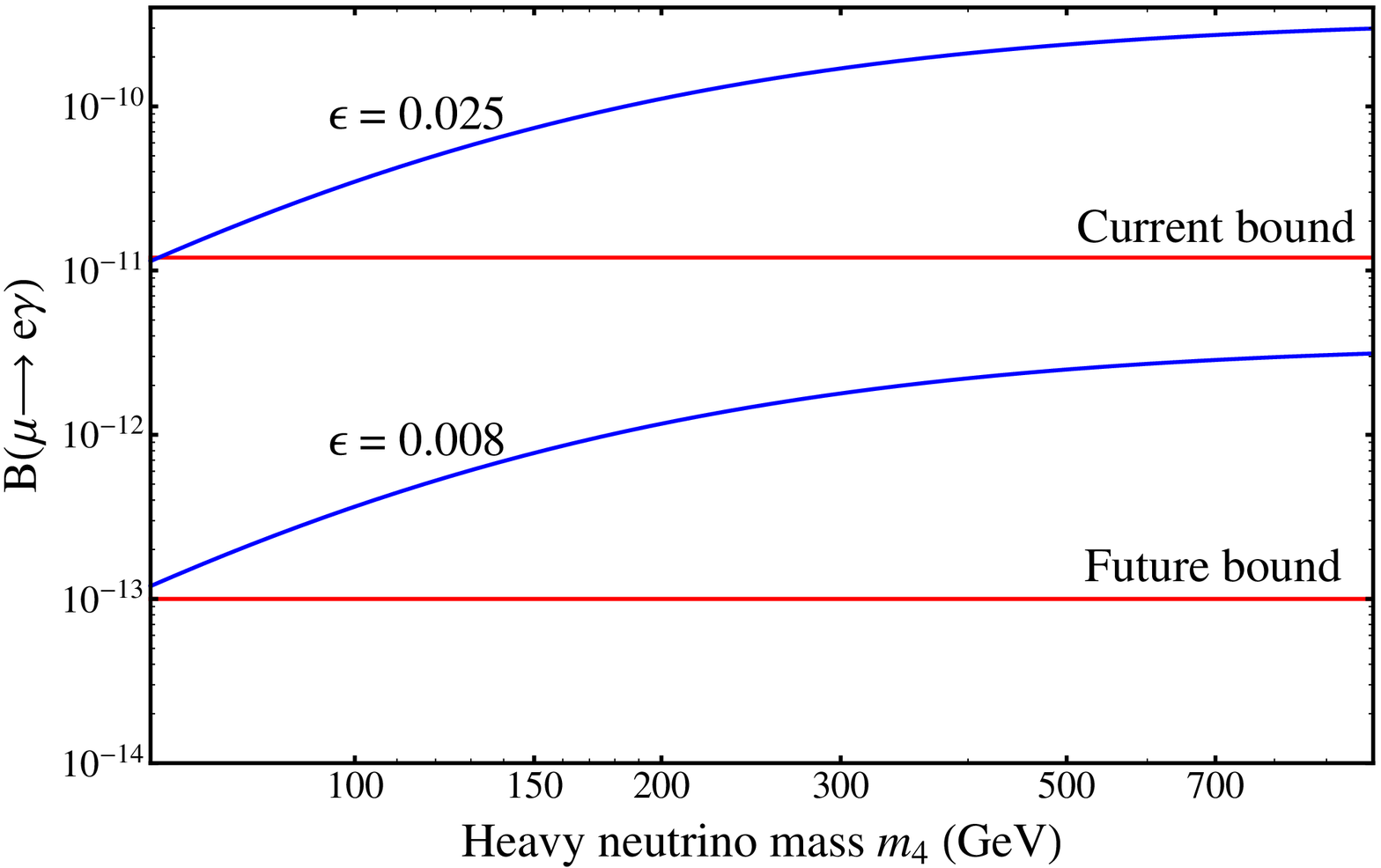, height=1.8in}
\qquad
\psfig{figure=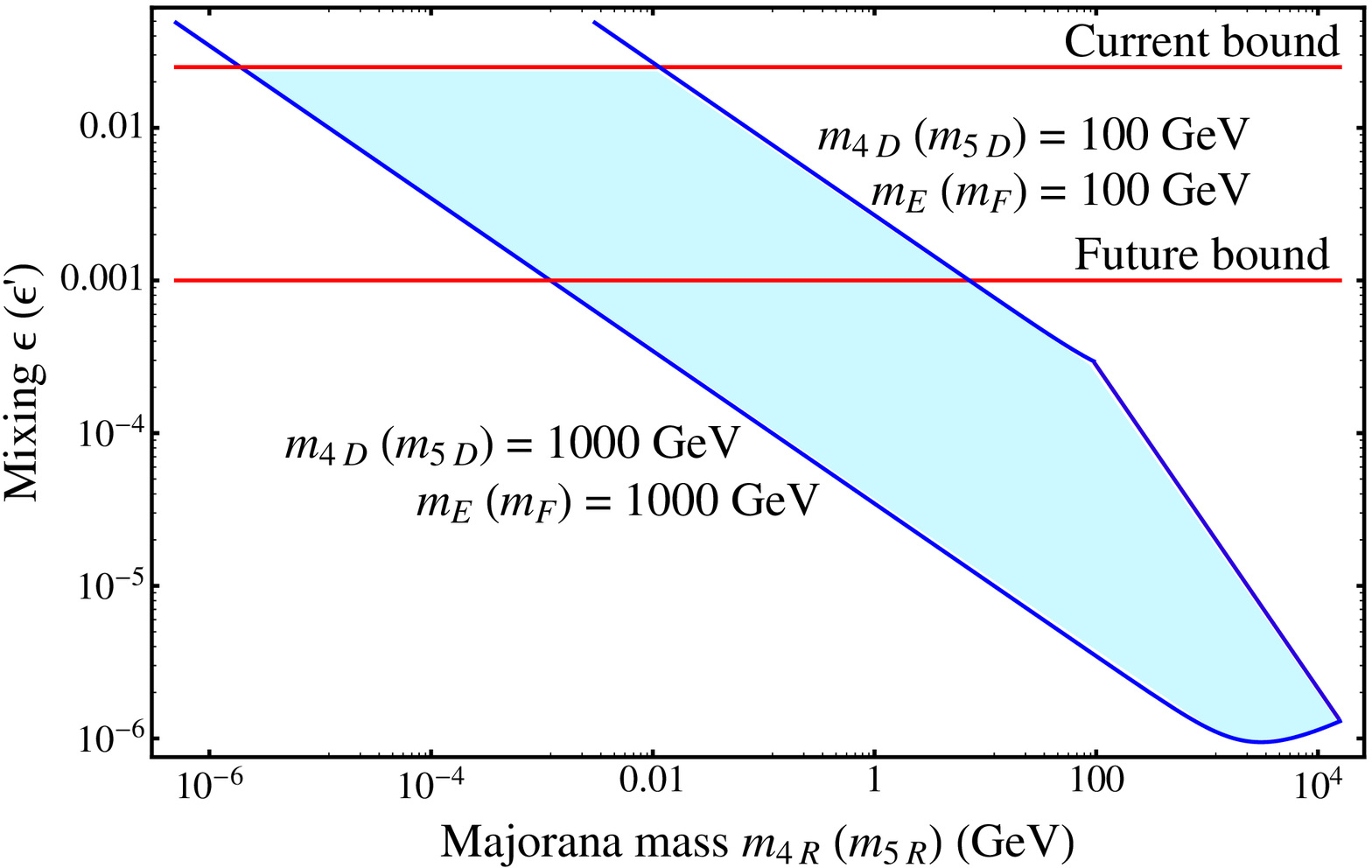, height=1.8in}
\caption{a) Left: $B(\mu\rightarrow e\gamma)$ against $m_{4}$ for different values of $\epsilon$. We also display present and future limits. \label{fig:mu2egamma}
b) Right: Parameter space that predicts the right scale for heavy and light neutrinos (the region between the curves). We also present the current $\mu\rightarrow e\gamma$ bound and the expected $\mu$--$e$ conversion limit. \label{fig:Parameter-space}}
\end{figure}

\section{Phenomenological analysis of the model \label{pheno}}

In general, the most restrictive experimental bound comes from $B(\mu\rightarrow e\gamma)<1.2\times10^{-11}$, and it is translated into $\epsilon<0.03$. From $B(\tau\rightarrow\mu\gamma)<4.4\times10^{-8}$, we obtain $|\epsilon^{\prime2}-\epsilon^{2}|<0.09$. We display in Figure 2 a) $B(\mu\rightarrow e\gamma)$ versus the mass of the heavy neutrino $m_{4}$ in the NH case. We also display present and near future limits. Also,  $\mu$--$e$ conversion in nuclei gives information on $\epsilon$. We expect it to set much stronger bounds in the future. 

Violations of universality constrain the model in both hierarchies. For example, from pion decay, we obtain $\epsilon'<0.04$. Regarding neutrinoless double beta decay, there are contributions of the new heavy neutrinos, however, we obtain the same combination of parameters as in the light neutrino masses expressions, $m_{4R}\epsilon^{2}$, leading to unobservable effects in $0\nu\beta\beta$ when the former are fitted (of course the light neutrino contribution can be observed (in IH) in the future).

Also, a very striking effect of new generations is the enhancement of the Higgs-gluon-gluon vertex which arises from a triangle diagram with all quarks running in the loop, by a factor of 9 (25) in the presence of a fourth (fifth) generation. We estimate roughly that $m_{H}>300\,\mathrm{GeV}$ in the case of five generations. However, these limits may be softened in some cases.

To summarize the phenomenology of the model we present in figure~\ref{fig:Parameter-space} the allowed regions in the $\epsilon-m_{4R}$ plane which lead to $m_{3}\sim0.05\,\mathrm{eV}$ varying the charged lepton masses $m_{E}\,(m_{F})$ and the Dirac neutrino masses $m_{4D}\,(m_{5D})$ between 100~GeV--1~TeV, and imposing the bound on the neutrino mass, $m_{4}>63\,\mathrm{GeV}$. We also plot the present bounds on the mixings $\epsilon\,(\epsilon^{\prime})$ from $\mu\rightarrow e\gamma$ and future limits from $\mu$--$e$ conversion if expectations are attained.

So our conclusion is that with four generations and one singlet the correct spectrum of light neutrino masses cannot be generated. However, with five generations and two singlets all current data can be accomodated in the region of the parameter space between the curves of Figure 2 b), which will be probed in the near future.

\section*{Acknowledgments}
This work has been partially supported by the grants FPA-2007-60323, FPA-2008-03373, CSD2007-00060, CSD2007- 00042, PROMETEO/2009/116, PROMETEO/2009/128 and MRTN-CT-2006-035482 (FLAVIAnet). A.A. and J.H.-G. are supported by the MICINN under the FPU program.

\section*{References}
\providecommand{\href}[2]{#2}\begingroup\raggedright\endgroup


\end{document}